\documentclass[prl,aps,twocolumn,superscriptaddress,showpacs]{revtex4}

\usepackage{epsfig}
\usepackage{amsmath}

\begin{document}

\title{
	Superfluid, solid, and supersolid phases of dipolar 
	bosons in a quasi-one-dimensional optical lattice 
}

\author{
	Jonathan M. Fellows
}\affiliation{
	School of Physics, 
	University of Birmingham, 
	Birmingham B15 2TT, 
	United Kingdom
}

\author{
	Sam T. Carr
}\affiliation{
	Institut f\"ur Theorie der Kondensierten Materie 
	and DFG Center for Functional Nanostructures, 
	Karlsruher Institut f\"ur Technologie, 
	76128 Karlsruhe, 
	Germany
}

\date{	\today	}


\begin{abstract}
We discuss a model of dipolar bosons trapped in a weakly coupled planar array of one-dimensional tubes.  We consider the situation where the dipolar moments are aligned by an external field, and find a rich phase diagram as a function of the angle of this field exhibiting quantum phase transitions between solid, superfluid and supersolid phases.
In the low energy limit, the model turns out to be identical to one describing quasi-one-dimensional superconductivity in condensed matter systems.  This opens the possibility of using bosons as a quantum analogue simulator of electronic systems, a scenario arising from the intricate relation between statistics and interactions in quasi-one-dimensional systems.
\end{abstract}

\maketitle

Experimental developments in ultracold atomic systems continue to yield insights into exotic quantum phases that were once the dominion of theorists. Since the first Bose-Einstein condensates were produced in the early nineties \cite{Anderson-etal-1995,Davis-etal-1995}, the steady march of experimental progress continues to allow us access to all manner of phases from Mott insulator to superfluids \cite{Greiner-etal-2002}  -- see Ref.~\onlinecite{Bloch-Dalibard-Zwerger-2008} for a recent review.  Solid theoretical proposals exist for many other unconventional collective states -- for example since early proposals to find the elusive supersolid in an optical lattice \cite{Scarola-DasSarma}, there has been a spate of recent suggestions \cite{Mathey-2009,Yamamoto-etal-2011,Pollet-etal-2010,He-Hofstetter-2011,Ohgoe-Suzuki-Kawashima-2011} of alternative setups in which a supersolid phase may be uncovered.

A promising possible application of these developments is that one might be able to use arrangements of ultracold atoms in optical traps as analogue computers for simulating strong  correlations in electronic condensed matter systems\cite{Quintanilla-Hooley}.  As correlations are enhanced by low dimensions, it becomes natural to consider systems involving weakly coupled  quasi-one-dimension tubes of particles, systems which share their underlying geometry with that of some of the more interesting condensed matter systems.  While it may at first seem preferable to deal with fermions if one intends to compare back to electronic systems, in this paper we will demonstrate a {\em bosonic} model with the same low-energy effective theory as a well studied model of striped superconductivity in electronic systems \cite{Efetov-Larkin,Carr-Tsvelik-2002} that has been successfully applied to the compound Sr$_{2}$Ca$_{12}$Cu$_{24}$O$_{41}$ \cite{Abbamonte-et-al-2004} but may even have relevance for high-$T_c$ cuprate materials \cite{HiTc-stripes}.  This equivalence between bosonic and fermionic models arises as a result of the quasi-one-dimensional nature of the setup, where the role of statistics and interactions are intrinsically intertwined \cite{Sutherland-Beautiful-Models} and implies that suitably chosen bosonic experiments can inform us about the low energy excitations of fermionic systems.

Experimental groups now routinely produce such tubes of bosons.  The obvious interaction to exploit within such systems is the contact interaction as this is tunable through Feshbach resonance \cite{Theis-etal-2004}. However contact interactions are too short-ranged to couple distinct tubes and so it becomes preferable to deal with the dipole-dipole interaction which itself is well tunable in these systems\cite{Stuhler-Pfau-2005} and allows the manufacture of custom spatially anisotropic interactions in such systems \cite{Huang-Wang,Goral-2002,Quintanilla-Carr-Betouras-2009,Fregoso-Sun-Fradkin-Lev-2009}. It is unimportant for the majority of our considerations whether the dipole moment is electric or magnetic, however one may desire the larger energy scales offered by the electric dipole to access some of the physics we will encounter.

\begin{figure}
	\begin{center}
		\includegraphics[width=2.8in,clip=true]{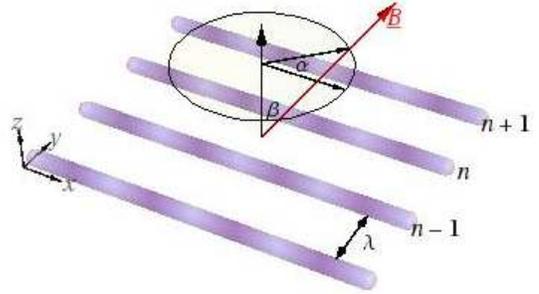}\vspace{-0.3cm}
	\end{center}
		\vspace{-5mm}
	\caption{[Color online] Proposed experimental set up.  An optical lattice in the $y$ direction separates the trapped dipolar bosons into weakly coupled tubes.  The dipole moments of the bosons are polarized by an external field, $\underline{B}$, which may be rotated in three dimensions to change the effective interactions.		\vspace{-0.6cm}}
	\label{Fig1}
\end{figure}

In this manuscript we will discuss such a model in which a gas of dipolar bosons is arranged in a two-dimensional planar array of one-dimensional homogeneous cigars -- see Fig.~\ref{Fig1}.  This may be created by a strong two dimensional optical lattice in which the confinement in the $z$-direction completely suppresses hopping in that direction, while that in the $y$-direction allows some weak hopping between the tubes. In principle, a weak optical lattice may also be present in the longitudinal ($x$) direction without changing the physics, so long as the lattice is incommensurate with the density of bosons. We allow for interactions between the tubes through a dipole-dipole interaction which can be tuned via the direction of the dipole moment.

A similar experimental set up has been considered in \cite{Kollath-Meyer-Giamarchi-2008} in the limit of zero inter-tube hopping.  The effects of this hopping, but in the absence of the dipole-dipole interaction was discussed in \cite{Ho-Cazalilla-Giamarchi-2004}.  In this paper, we will consider both inter-tube couplings to be present, paying particular attention to the quantum critical region created by their competition. Given the anisotropy of the lattice and the freedom of the dipole moment to point anywhere in 3-space, we have a great control over the interaction strength in the longitudinal and transverse directions. If we choose the angle $\beta$ between the dipole-moment and the normal to the plane to satisfy $\sin\beta =1/\sqrt{3}$ ($\beta$ is referred to in the literature \cite{Quintanilla-Carr-Betouras-2009} as the \emph{magic angle}) then interactions in either direction will be repulsive and it is this case that we will consider from here on, fixing $\beta$ and varying $\alpha$.

The dipole-dipole interaction may be split into two contributions: intra-tube and inter-tube.   While we keep explicitly the full structure in the intra-tube part, we approximate the inter-tube part by a local nearest-neighbor interaction, this approximation is valid for the dipolar  interaction when, as is the case here, the anisotropy of the underlying lattice is large. For a fuller discussion see \cite{Quintanilla-Carr-Betouras-2009}, note here though that the dipole interaction in two dimensions (as opposed to three dimensions) is not truly long ranged as there is no divergence in the integral of the interaction. This means that none of the essential physics is changed when considering nearest neighbour interactions only as we are regardless operating within the correct universality class. Now if $x$ measures along tube $n$ (at $y=n\lambda$)  and $\alpha$ is the angle between the horizontal projection of the dipole moment and the axis of the tubes (see Fig.~\ref{Fig1}), then the contributions to the dipole-dipole interaction take the form

\vspace{-5mm}

\begin{align}
	&V_\text{dd}^\text{intra} = \frac{C_\text{dd}}{4\pi} \frac{\sin^2\alpha}{|x-x'|^3} \delta_{nn'},\nonumber\\
	&V_\text{dd}^\text{inter}  =\frac{C_\text{dd}}{4\pi} \frac{\cos^2\alpha}{\lambda^3}\delta(x-x')\delta_{n,n'\pm 1},
\end{align}

\noindent where $C_\text{dd}$ is the strength of the dipole-dipole interaction.

The system is therefore described by a Hamiltonian $\hat{H} = \sum_n \hat{H}^{(n)}_\text{tube}+\hat{H}^{(n)}_\text{inter}$, where the individual contributions are given by

\vspace{-7mm}

\begin{align}
	&\hat{H}^{(n)}_\text{tube} = \int dx~\hat{\Psi}_n^\dag(x) \left(-\frac{\hbar^2}{2m} \partial_x^2\right) \hat{\Psi}_n(x)  \nonumber\\
	&\qquad+ \int dx\,dx'\, \hat{\rho}_n(x)  V_\text{dd}^\text{intra}(x,x')  \hat{\rho}_n(x') ,\nonumber\\
	&\hat{H}^{(n)}_\text{inter} = -\int dx\,   J_\perp\left( \hat{\Psi}_n^\dag(x)  \hat{\Psi}_{n+1}(x) + \hat{\Psi}_{n+1}^\dag(x)  \hat{\Psi}_n(x) \right) \nonumber\\
	&\qquad+ \int dx\,dx'\, \hat{\rho}_n(x) w V_\text{dd}^\text{inter}(x,x') \hat{\rho}_{n+1}(x').
\end{align}

\noindent Here,  $\hat{\Psi}^\dag_n(x)$ denotes the creation field operator in the $n^\text{th}$ tube, the hopping parameter $J_\perp$ may be found by integrating the confining potential with respect to the appropriate Wannier functions of the longitudinal lattice, and $w$ comes from the integral off the Wannier functions in adjacent chains.  We may also include a contact interaction in the above model, however for simplicity we consider the case where this is negligible, as in \cite{Lahaye-Pfau-2007}.

The above described model has two clear strong coupling limits; when the interaction is dominant, the ground state will be a density wave (DW) while if interactions are weak a superfluid (SF) phase should emerge.   We will now derive the phase diagram of this model as a function of experimentally controllable parameters, paying particular attention to the quantum critical region between the DW and SF, where we show that a supersolid phase may also manifest itself.

Firstly we bosonize the Hamiltonian according to the standard procedure \cite{Cazalilla-2011,Giamarchi-2004} via the prescription

\begin{align}
	\hat{\Psi}^\dag_n(x)&\approx\sqrt{A\rho_0}e^{-i\hat{\theta}_n(x)}, \\
	\hat{\rho}_n(x)&\approx\rho_0-\frac{1}{\pi}\partial_x\hat{\varphi}_n(x)+2\rho_0\cos[2\pi\rho_0 x - 2 \hat{\varphi}_n(x)], \nonumber
\end{align}

\noindent where $\rho_0$ is the particle density (which we assume to be the same for each tube), $A$ is a non-universal dimensionless constant  of order one and $\hat{\varphi}_n$ and $\hat{\theta}_n$ are the dual fields of the $n^\text{th}$ tube ($[\hat{\varphi}(x),\partial_x\hat{\theta}(x')]=i\delta(x-x')$).

Now neglecting higher harmonics, additive constants and terms which vanish faster than $1/\rho_0$ we arrive at the following bosonized description

\begin{align}
	&\hat{H}^{(n)}_\text{tube} = \frac{\hbar v_s}{2\pi}\int dx\left[K\left(\partial_x\hat{\theta}_n\right)^2+\frac{1}{K}\left(\partial_x\hat{\varphi}_n\right)^2 \right],\nonumber\\
	&\hat{H}^{(n)}_\text{pair} = \frac{\hbar v_s}{2\pi}\int \frac{dx}{a^2} \Big[ g_\text{int} \cos2\left(\hat{\varphi}_n-\hat{\varphi}_{n+1}\right)  \nonumber\\ 
	&\qquad\qquad\qquad - g_\text{hop} \cos\left(\hat{\theta}_n-\hat{\theta}_{n+1}\right) \Big].
\end{align}

\noindent Here $a\sim\rho_0^{-1}$ is the short distance cutoff and the dimensionless coupling constants $g_\text{int}$ and $g_\text{hop}$ are given by

\begin{equation}
	g_\text{int} = \frac{a^2\rho_0 w C_\text{dd} \cos^2\alpha}{\lambda^3\hbar v_s}, \;\;\; g_\text{hop} = \frac{4 \pi a^2 A J_\perp \rho_0}{\hbar v_s}.
\end{equation}

\begin{figure}
	\begin{center}
		\includegraphics[width=2.8in,clip=true]{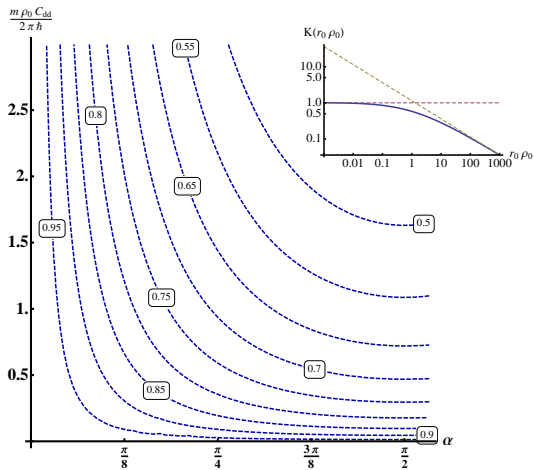}\vspace{-0.3cm}
	\end{center}
	\caption{[Color online] A contour plot showing the variation of the Luttinger parameter $K$ with the dipole angle $\alpha$ and the relative strength of the dipole interaction. The most important contour for our purposes is that for $K=\frac{1}{2}$ as we shall see this is where the DW and SF phases are maximally in competition. Inset: the bare relation between $K$ and dipole strength based on the parameterization in Ref.~\onlinecite{Citro-etal-2008}. Although for cold atoms with a magnetic dipole moment, $\rho_0r_0\ll 1$, cold molecules with an electric dipole can realistically be in the regime where $\rho_0 r_0>1$\cite{Citro-etal-2007}. 
\vspace{-0.5cm} }
	\label{Fig2}
\end{figure}

In the absence of inter-tube coupling ($g_\text{hop}=g_\text{int}=0$), each tube is governed by a Luttinger liquid type Hamiltonian with Luttinger exponent $K$ and sound speed $v_s$.  The relationship between these parameters and the original model parameters is non-universal, but has been determined numerically \cite{Citro-etal-2008}, where it is found that $K$ varies smoothly as a function of $\rho_0r_0$ where $r_0$ is the effective Bohr radius.   In our case, $r_0$ is given in terms of the dipole angle $\alpha$ by $r_0(\alpha) = \frac{m C_\text{dd}}{2\pi\hbar^2}\sin^2\alpha$, giving the dependence of $K$ on $\alpha$ as shown in Fig.~\ref{Fig2}.

This bosonized Hamiltonian is now of a form similar to that of a quasi-one-dimensional superconductor, so we shall adopt the approach taken in Ref.~\onlinecite{Carr-Tsvelik-2002}. If the interaction coefficient or the hopping coefficient is clearly dominant then in either limit we can consider the mean field behavior and arrive at a sine-Gordon model that must be solved along with a self-consistency condition. Setting $\beta_\text{int}=2K^{1/2}$ and $\beta_\text{hop} = K^{-1/2}$,  we find a mean field Hamiltonian of the form

\vspace{-3mm}

\begin{align}
	 &\hat{H}^\text{mf} = \frac{\hbar v_s}{2\pi}\int dx\left[ (\partial_x\hat{\theta} )^2+\left(\partial_x\hat{\varphi}\right)^2 +\frac{g^\text{mf}}{a^2} \cos \left(\beta\hat{\varphi} \right) \right], \nonumber\\
	&g^\text{mf} = 2 g \langle \cos(\beta\varphi) \rangle.
\end{align}

Though we explicitly write this for the interaction dominated case, due to the duality between $\hat{\theta}$ and $\hat{\varphi}$ we can treat both regimes simultaneously.

The expectation value $\langle \cos(\beta\hat{\varphi}) \rangle$ is known within the sine-Gordon model \cite{Lukyanov-Zamolodchikov-1997} so we can solve the self consistency conditions to find that $g^\text{mf}\sim 2 g^{\frac{1}{2}\left(1+\frac{1}{1-d}\right)}$, where $d=\beta^2/4$ is the scaling dimension of the operator in question. The scaling dimensions of our two inter-chain perturbations are $d_\text{int} = K$ and $d_\text{hop} = 1/(4K)$.

 Inspecting the mean field solutions we have acquired, we see that the hopping dominated region is characterized by a non-zero expectation of $\langle\cos(\hat{\theta})\rangle$ which corresponds to  $\langle \Psi_n^\dagger (x)  \rangle \ne 0$, and as such represents a SF phase.  In the opposite case when the interaction dominates, the phase is characterized by a non-zero expectation of $\langle\cos(2\hat{\varphi})\rangle$ which corresponds to  $\langle\rho_n(x) \rangle = \rho_0 + C (-1)^n\cos(2\pi\rho_0x)$,  and hence represents a checkerboard DW phase. Notice that if we were to abandon the ``magic angle'' which is holding all interactions repulsive and allow for attractive interactions, the DW phase can also be driven into a striped phase, a detailed study of the phase boundary of these two DW phases is undertaken in  \cite{Kollath-Meyer-Giamarchi-2008}. By changing $\beta$ one could see either DW phase competing with the SF phase, for concreteness we will hereon continue to assume that the relevant DW phase is the checkerboard.

In order to determine which phase is dominant for a given parameter range, and hence find the phase diagram, we consider both the hopping and interaction terms as perturbations over a Luttinger liquid and ask which has the higher critical temperature. We find the SF and DW susceptibilities ($\chi_\text{sf}$ and $\chi_\text{dw}$ respectively) within the random phase approximation using $\chi = \chi^{(0)}/(1-\frac{g v_s}{a^2}\chi^{(0)})$. The LL susceptibility for either phase is (see e.g. \cite{Giamarchi-2004})

\begin{equation}
	\chi^{(0)}\sim\frac{1}{2}\frac{a^2}{v_s}\left(\frac{\pi a k_B T}{\hbar v_s}\right)^{2d-2}.
\end{equation}

The critical temperature $T_c$ is estimated as the temperature at which an instability arises from a divergence in the associated susceptibility.  As our system is two-dimensional, these transitions are in the Berezinskii-Kosterlitz-Thouless universality class, however this simple calculation still estimates $T_c$ correctly \cite{Carr-Tsvelik-2002} in the weak coupling regime.  Hence we find $T_c$ from the condition  $\chi^{(0)}\sim\frac{a^2}{gv_s}$ so we find that the interactions dominate and the system takes on a DW phase when $g_\text{int}^\frac{1}{2-2K}\gg g_\text{hop}^\frac{2K}{4K-1}$. In the opposing region the inter-site hopping dominates and the system becomes a superfluid.

\begin{figure}
	\begin{center}
		\includegraphics[width=3.15in,clip=true]{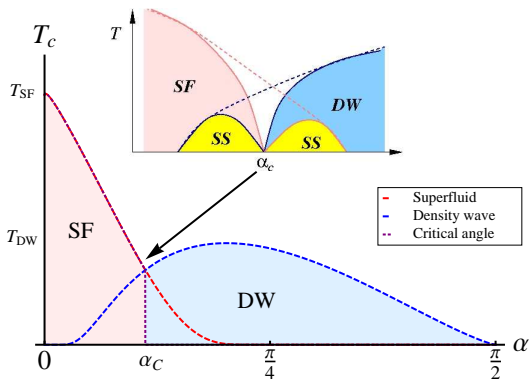} \vspace{-0.3cm}
	\end{center} \vspace{-10mm}
	\caption{[Color online] A plot of the critical temperature for the DW and SF phases and the implied phase diagram for an illustrative set of parameters ($g_\text{int}=0.1$, $g_\text{hop}=0.2$ and $R=250$) chosen such that $K\sim\frac{1}{2}$ at the critical point.  Inset: correction the phase diagram at the critical angle showing the onset of a supersolid (SS) phase. Based on realistic experimental parameters for the dipole moment of e.g. KRb \cite{Lahaye-et-al} and a lattice spacing $\lambda\sim 500nm$, we estimate the maximum critical temperature of the DW phase to be of order $T_{DW}\sim 50nK$.}
	\label{Fig3}
\end{figure}

The parameter we wish to focus upon is the angle $\alpha$ which comes from the geometry of our proposed setup. The explicit dependence of the critical temperature of the SF and DW phases on this geometric parameter are given by

\begin{align}
	&k_B T_c^\text{SF} = \frac{\hbar v_s}{\pi a}\left(\tfrac{1}{2}g_\text{hop}\right)^\frac{2K(Rg_\text{int}^\text{max}\sin^2\alpha)}{4K(Rg_\text{int}^\text{max}\sin^2\alpha)-1},\nonumber\\
	&k_BT_c^\text{DW} = \frac{\hbar v_s}{\pi a}\left(\tfrac{1}{2}g_\text{int}^\text{max}\cos^2\alpha\right)^\frac{1}{2-2K(Rg_\text{int}^\text{max}\sin^2\alpha)},
\end{align}

\noindent where $R=\frac{m\lambda^3 v_s}{2\pi\hbar w a^2\rho_0}$ and $g_\text{int}^\text{max}={\rm max}(g_\text{int}(\alpha))=g_\text{int}(\alpha=0)$. Fig.~\ref{Fig3} shows the dependence of these critical temperatures on $\alpha$ for indicative values of  $R$, $g_\text{hop}$ and $g_\text{int}^\text{max}$.

The above quasi-one-dimensional analysis is valid as long as the parameters are sufficiently far from criticality.  In fact, even though the ordered phases are necessarily two-dimensional, deep in either of them, the correlation functions retain their one-dimensional character \cite{Carr-Tsvelik-2003}.  However, near the quantum-critical point (QCP) between the DW and SF, the entire quasi-one-dimensional formalism breaks down, and a full two-dimensional field theory is needed, the dimensional anisotropy here being an irrelevant operator \cite{Jaefari-Lal-Fradkin-2010}.

Exactly at the QCP, where DW can be smoothly rotated into SF, the order parameter acquires an enlarged (and non-Abelian) symmetry \cite{Carr-Tsvelik-2002}. This is in analogy with $SO(5)$ theories of the antiferromagnetic-superconductor transition\cite{Demler-Hanke-Zhang}, except that  the enhanced symmetry at the SF-DW transition is O(4)\cite{Jaefari-Lal-Fradkin-2010}. Following Jaefari {\em et al.}~\cite{Jaefari-Lal-Fradkin-2010}, in the vicinity of the QCP we can then write a $2+1$ dimensional non-linear sigma model in terms of this enhanced $O(4)$ order parameter, which has a Lagrangian

\vspace{-3mm}

\begin{equation}
	{\cal L}_\text{eff}[N] = \frac{1}{2g_0} (\partial_\mu N)^2 - w \partial_\mu N_a {\cal O}_{ab} \partial_\mu N_b - \tilde{h} N_a \tilde{\cal O}_{ab} N_b.
\end{equation}

\noindent Here, $N$ is the order parameter, $g_0\sim \sqrt{g_\text{int} + g_\text{hop}}$ is the effective coupling constant, $w\sim K-\frac{1}{2}$ signifies the departure from the point where the two inter-chain terms have the same scaling dimension, $\tilde{h}\sim g_\text{int}-g_\text{hop}$ measures the distance from the critical point, and ${\cal O}$ and $\tilde{\cal O}$ are two fixed matrices that break the enhanced symmetry in the appropriate way.  It turns out \cite{Jaefari-Lal-Fradkin-2010} that the term proportional to $w$ is marginally irrelevant in the renormalization group (RG) sense, meaning that the generic quantum critical behavior of our original model may be described by the above non-linear sigma model.  Note that all we are doing here is writing the most general Langrangian with all the correct symmetries. This model has been well studied using RG methods \cite{Calabrese,Jaefari-Lal-Fradkin-2010} where it is found that there is a tetracritical point at $g_0=g_c \sim 1$ separating different types of critical behavior.  In the case $g_0<g_c$ which is always satisfied in our present weakly-coupled chains scenario, there is a region around $\alpha_c$ where there is coexistence of both the DW and SF order parameters, in other words a super-solid phase.  The corrected phase diagram in the vicinity of the critical point is shown schematically in the inset to Fig.~\ref{Fig3}.

Unfortunately precise predictions regarding the supersolid phase are not easily attainable because this phase is deep within the strongly correlated regime and not accessible from our interchain perturbation theories. Its existence must rather be inferred based on RG relating to the symmetries of the order parameter at the quantum critical point, leading to a qualitative phase diagram. This said, as indicated in to Fig.~\ref{Fig3}, the  relevant region to look for the supersolid phase in any experimental investigations of this system is set by the transition temperature of the suppressed phase, which could be found by our mean field calculations.

We can compare our results to those found in isotropic 2D analogs of this system: in an isotropic lattice with nearest neighbour interactions it has been shown that \cite{Scarola-DasSarma} the supersolid phase may arise at the interface of the SF and DW phases. The stability of the SS phase is considered in \cite{Sengupta} in which they note that this phase is possible in the presence of a lattice.  This is as opposed to the free system in which, for a dipole-dipole interaction, it has been seen \cite{Filinov} that there is a depletion of the SF phase before the onset of a dipole solid phase. Although our method is not applicable to the isotropic limit, the fact that the phase diagrams for models on a 2D lattice are so similar suggests that there will not be any additional transitions as we decrease the anisotropy.

Returning finally to the proposed experimental setup in which we expect this phase diagram to manifest itself; some groups \cite{Sherson-et-al-2010} have reported \emph{single-site} resolution in flourescence imaging of $^{87}\text{Rb}$ in a two dimensional lattice setup similar to that we describe so it may well be possible to probe the DW phase of this system \emph{in situ}.   A common way to probe the SF phase is releasing the bosons from their confinement, and looking for high contrast in the interference pattern in the absorbtion data \cite{Greiner-etal-2002}.  The hallmark of the supersolid phase we have described would be the presence of both a DW and SF order parameter.

In summary, we have shown how a popular model in condensed matter physics of striped superconductivity may also appear in a cold atom experiment with trapped dipolar {\em bosons} in a quasi-one-dimensional geometry.  The rich phase diagram of the model may be easily explored by changing the angle of a polarizing field, and shows crystalline, superfluid and supersolid phases, on top of the high-temperature sliding Luttinger liquid phase.

We  thank Joseph Betouras, Dima Gangardt, Chris Hooley and Jorge Quintanilla for useful discussions.


\end{document}